\documentclass[11pt]{article}

\usepackage{url,amsmath,amsthm,amscd,amsfonts,graphicx,psfrag}
\usepackage[all]{xy}

\let\0=\over
\def\2{{1\over2}}
\let\6=\partial
\def\({\left(}       \def\){\right)}

\let\bra=\langle        \let\ket=\rangle        \def\<#1\>{\bra #1 \ket}

\def\rd{{\rm d}}
\def\Tr{{\rm Tr}}
\def\hX{{\widehat X}}
\def\hhX{{[\widehat X]}}
\def\hpi{{\widehat \pi}}
\def\hH{{[\widehat H]}}
\def\hA{{[\widehat A]}}
\def\he{{[\widehat e]}}
\def\hl{{[\widehat \ell]}}
\def\qX{{[X]}}
\def\qH{{[H]}}
\def\qA{{[A]}}
\def\qe{{[e]}}

\def\qE{{[E]}}
\def\qpt{{[\text{pt}]}}
\def\qHA{{[H_A]}}
\def\Eh{{[E_1]}}
\def\Ev{{[E_2]}}
\def\Ed{{[E_d]}}
\def\Pic{{\rm Pic }}
\def\che{{\rm ch}}
\def\td{{\rm td}}

\newcommand{\be}{\begin{equation}}
\newcommand{\ee}{\end{equation}}

   \def\cl{{\cal L}}

\def\IR{{\mathbb R}}  \def\IP{{\mathbb P}}
   
\def\IZ{{\mathbb Z}} \def\IQ{{\mathbb Q}}

\theoremstyle{plain}
\newtheorem{thm}{Theorem}

\newtheorem{lem}[thm]{Lemma}

\theoremstyle{remark}
\newtheorem*{rem}{Remark}

\theoremstyle{definition}
\newtheorem{defn}{Definition}
\newtheorem*{TWC}{The weak heterotic challenge}
\newtheorem*{TSC}{The strong heterotic challenge}

\numberwithin{equation}{section}

\usepackage[scale=0.70]{geometry}

\usepackage{sectsty}
\sectionfont{\center \sc \large}
\subsectionfont{\bf \large}
\subsubsectionfont{\it \large}

\usepackage{setspace}

\begin{document}
\begin{titlepage}

{\noindent \Large \bf \fontfamily{pag}\selectfont Exploring a new peak in the heterotic landscape}

\vspace*{0.5cm}
\noindent \rule{\linewidth}{0.5mm}

\vspace*{1.3cm}

{\noindent  \fontfamily{pag}\selectfont  Anthony Bak}\\[0.2em]
{\small {\it \indent Stanford University and American Institute of Mathematics}\\[0.1em]
\indent \url{bak@math.stanford.edu}}\\[0.3em]

{\noindent \fontfamily{pag}\selectfont  Vincent Bouchard}\\[0.2em]
{\small {\it \indent Department of Mathematical and Statistical Sciences\\
\indent University of Alberta}\\[0.1em]
\indent \url{vincent@math.ualberta.ca}}\\[0.3em]

{\noindent \fontfamily{pag}\selectfont  Ron Donagi}\\[0.2em]
{\small {\it \indent Department of Mathematics\\
\indent University of Pennsylvania}\\[0.1em]
\indent \url{donagi@math.upenn.edu}}\\

\vspace*{1.5cm}

\hspace*{1cm}
\parbox{11.5cm}{{\sc Abstract:} We study the existence of realistic heterotic vacua on a new Abelian surface fibered Calabi-Yau threefold $X$ with $\IZ_8 \times \IZ_8$ fundamental group. Our main result is a no-go theorem, which says that (under mild assumptions) there is no stable holomorphic vector bundle on $X$ satisfying the constraints required by global consistency of the heterotic vacuum and phenomenology. To prove the theorem we explore in some detail the Fourier-Mukai transform of vector bundles on Abelian surface fibrations.}

\end{titlepage}

\setstretch{1.08}
\tableofcontents 

\section{Introduction}

As a further step in the search for phenomenologically realistic compactifications of string theory, we explore compactifications of the $E_8 \times E_8$ heterotic string on a new Calabi-Yau threefold $X$ with fundamental group $\pi_1(X) = \IZ_8 \times \IZ_8$.

The Calabi-Yau threefold was constructed a few years ago by Gross and Popescu \cite{GP}. It has been singled out various times recently as a promising candidate on which to compactify the heterotic string, due to its large fundamental group --- it is the Calabi-Yau threefold with the largest fundamental group known to date. Indeed, the most successful method for implementing GUT breaking in the heterotic string consists in using discrete Wilson lines, which require that the compactification space be non-simply connected. The existence of a fundamental group as large as $\IZ_8 \times \IZ_8$ seems very promising, since it opens up the possibility for various symmetry breaking mechanisms.

In this paper we prove a general no-go theorem which denies (under mild assumptions) the existence of realistic heterotic compactifications on this Calabi-Yau threefold. The proof of the no-go theorem exhibits an explicit tension between a defining condition for global consistency of heterotic vacua, namely anomaly cancellation, and the phenomenological requirement of a three-generation low-energy theory. This tension parallels the analysis presented in \cite{BD2} in the context of heterotic compactifications on another non-simply connected Calabi-Yau threefold. In some way, for compactifications on Gross-Popescu's threefold, it turns out that requiring \emph{global consistency}, or UV-completion, of the heterotic vacuum throws out all ``local vacua" satisfying the numerical constraints required by phenomenology.

In the low-energy effective field theory limit, supersymmetric heterotic vacua are defined by a Calabi-Yau threefold $X$ and a stable holomorphic vector bundle on $X$ satisfying the anomaly cancellation constraint. In the much studied context of heterotic compactifications on elliptically fibered Calabi-Yau threefolds, a very useful approach to the construction of stable holomorphic vector bundles consists in using the so-called Fourier-Mukai transform, also known as the spectral cover construction \cite{FMW,FMW2}. The Fourier-Mukai transform provides a dual description which in many instances simplifies the construction of stable vector bundles. 

However, Gross-Popescu's threefold does not admit an elliptic fibration; rather, it possesses a fibration over $\IP^1$ with Abelian surface ($T^4$) fibers. The Fourier-Mukai transform on Abelian surface fibered Calabi-Yau threefolds is studied in detail in the companion mathematical paper \cite{Ba} by the first author. We make use of this mathematical construction to study the landscape of stable bundles on Gross-Popescu's threefold and prove our main no-go theorem.

While the proof of the no-go theorem is now rather clean, its historical origins included many twists and turns which may be worth recounting, since not only are they somehow amusing, but they also expose the inherent complexity of constructing globally consistent heterotic vacua. Starting from the fact that there is still only one known globally consistent compactification of the heterotic string with precisely the MSSM massless spectrum \cite{BD1,BCD}, our initial goal was to use Gross-Popescu's threefold to construct other realistic compactifications of the heterotic string. So we approached the problem with hope and determination, and tried to construct stable vector bundles satisfying the phenomenological and global consistency constraints using all the algebro-geometric techniques at our disposal. We managed to translate the construction into some algorithmic procedure that could be handled numerically, and literally produced hundreds of thousands of candidate ``bundles" satisfying the numerical constraints. At this stage, we were very confident that at least one of these hundreds of thousands of candidates would provide a globally consistent heterotic vacuum.

But we then started examining these models on a case-by-case basis, starting with those that seemed the most promising to us. Only to find out, time after time, that for each of the models that we studied, various aspects of the construction, such as stability of the bundle, anomaly cancellation, or surjectivity of a given map, were ill-defined. After a while, we realized that perhaps in fact there was a fundamental reason why none of these models worked. But it was far from obvious, \emph{a priori}, that a no-go theorem was pulling the strings behind the scene --- indeed, the no-go theorem seems to involve an intricate interplay between various requirements of heterotic compactifications. Clearly, the search for realistic bundles in heterotic string theory would greatly benefit from a better algorithmic or intuitive understanding.

Here is an outline of the paper. In section 2, we review generalities of compactifications of the $E_8 \times E_8$ heterotic string. In section 3 we introduce Gross-Popescu's threefold, on which we compactify the heterotic string. We describe the geometry in some detail, and study the cohomology ring of the threefold. Our main no-go theorem concerning the existence of realistic bundles on Gross-Popescu's threefold is presented in section 4, which also contains the proof. Section 4.1 describes the Fourier-Mukai transform of vector bundles on Gross-Popescu's threefold. The proof of the theorem is then divided in two parts. In section 4.2, we use the results of section 4.1 to show that the three-generation condition implies that realistic vector bundles cannot be semistable on all Abelian fibers. However, we show that they can be ``transformed" into auxiliary bundles which are semistable on all fibers through a finite sequence of elementary transformations. Then in section 4.3 we show that the existence of this sequence of elementary transformations contradicts the phenomenological and global consistency requirements on the Chern character of our physical bundle, completing the proof of the no-go theorem.

\subsection*{Acknowledgments}
We would like to thank Mark Gross and Tony Pantev for many useful discussions and collaboration at the initial stages of this project. We would also like to thank Rhys Davies for useful correspondence. The research of R.D. is supported by NSF grant DMS 0612992 and Research and Training Grant DMS 0636606. The work of V.B. is supported in part by the Center for the Fundamental Laws of Nature at Harvard University and by NSF grants PHY-0244821 and DMS-0244464.


\section{Heterotic generalities}

\subsection{Heterotic vacua}

We consider the low-energy effective field theory limit of the $E_8 \times E_8$ heterotic string, which is given by 10-dimensional $N=1$ supergravity  coupled to $E_8 \times E_8$ super Yang-Mills theory. To obtain an effective 4-dimensional theory, we compactify on a 10-dimensional product space $M \times X$, where $M$ is 4-dimensional Minkowski space and $X$ is an internal 6-dimensional smooth compact manifold. We require unbroken supersymmetry in 4 dimensions, which, under simplifying assumptions (see for instance \cite{GSW2}), implies that $X$ must be a Calabi-Yau threefold and that the Yang-Mills gauge field strength $F$ must satisfy the Hermitian Yang-Mills equations. Furthermore, the condition
\be\label{e:anom1}
\Tr R^2 - \Tr F^2 = \rd H
\ee
is required for the anomaly to cancel, where $R$ is the Ricci tensor and $H$ is the field strength of the $B$-field.

Such compactifications can be reformulated in the language of algebraic geometry. First, the Yang-Mills gauge field corresponds to a connection on some vector bundle $V \to X$ on the Calabi-Yau threefold $X$. A well-known theorem of Donaldson-Uhlenbeck-Yau states that given a holomorphic vector bundle $V \to X$ and a K\"ahler class $\omega$ on $X$, there exists a unique connection satisfying the Hermitian Yang-Mills equations if and only if $V$ is \emph{polystable} with respect to $\omega$. 

To define polystability, recall that the slope $\mu_\omega (V)$ of a vector bundle $V$ with respect to an ample class $\omega$ on a manifold $X$ of dimension $d$ is given by
\be
\mu_\omega (V) = \frac{c_1(V) \cdot \omega^{d-1}}{\text{rank}(V) },
\ee
where $c_1(V)$ is the first Chern class of $V$. A vector bundle $V$ is \emph{stable} (resp. \emph{semistable}) with respect to $\omega$ if, for all coherent subsheaves $V'$ of $V$ with $0 < \text{rank}(V') < \text{rank}(V)$, we have $\mu_\omega(V')<\mu_\omega(V)$ (resp. $\mu_\omega(V') \leq \mu_\omega(V)$). It is \emph{polystable} if it is the direct sum of stable vector bundles of equal slope.

Moreover, \eqref{e:anom1} has a solution only if the topological condition
\be\label{e:anom}
c_2(X) - c_2(V) = 0
\ee
is satisfied, where $c_2(X)$ is the second Chern class of the tangent bundle of $X$, and $c_2(V)$ is the second Chern class of $V$.

We can summarize the discussion in the following definition:

\begin{defn}
A \emph{weakly coupled SUSY heterotic vacuum} is given by a triple $(X,\omega,V)$, where $X$ is a Calabi-Yau threefold, $\omega$ is a K\"ahler class on $X$, and $V \to X$ is a holomorphic vector bundle on $X$ with structure group $H \subseteq E_8 \times E_8$, satisfying the conditions:
\begin{itemize}
\item $V$ is polystable with respect to $\omega$;
\item $c_2(X) - c_2(V) = 0$.
\end{itemize}
\end{defn}

Note that the anomaly cancellation condition can be relaxed slightly. Following Ho\v rava and Witten \cite{HW,HW2}, it is believed that the strong coupling limit of heterotic string theory is $11$-dimensional M-theory compactified on the interval $S^1 / \IZ_2$. However, heterotic M-theory vacua can also include $M5$-branes wrapping holomorphic curves in $X$, and located at points on the interval. Those contribute to the anomaly, and the anomaly cancellation condition becomes
\be
c_2(TX) - c_2(V) = [M]
\ee
in the presence of an $M5$-brane, where $[M]$ is the effective cohomology class Poincar\'e dual to the holomorphic curve wrapped by the $M5$-brane. Therefore, if one is willing to consider the strong coupling limit of heterotic string theory, we obtain:

\begin{defn}
A \emph{strongly coupled SUSY heterotic vacuum} is given by a triple $(X,\omega,V)$, where $X$ is a Calabi-Yau threefold, $\omega$ is a K\"ahler class on $X$, and $V \to X$ is a holomorphic vector bundle on $X$ with structure group $H \subseteq E_8 \times E_8$, satisfying the conditions:
\begin{itemize}
\item $V$ is polystable with respect to $\omega$;
\item $c_2(X) - c_2(V)$ is a non-zero effective class.
\end{itemize}
\end{defn}

\subsection{Phenomenology}

Once we are given a weakly or strongly coupled heterotic vacuum $(X,\omega,V)$, the next step is to study phenomenology of the low energy four-dimensional theory.

\subsubsection{Gauge group}

The $E_8 \times E_8$ 10-dimensional gauge group is broken to the subgroup that commutes with the gauge fields present in the vacuum; that is, the resulting 4-dimensional gauge group is given by the commutant $G$ of the structure group $H \subseteq E_8 \times E_8$ of the vector bundle $V$. In particular, we will focus on the two following cases:
\begin{itemize}
\item $H=SU(5)$, which gives $G = SU(5) \times E_8$;
\item $H=SU(4)$, which gives $G = SO(10) \times E_8$.
\end{itemize}
Those two choices provide interesting candidates for Grand Unified Theories (GUT), with an extra hidden $E_8$ gauge theory. In the following we will use $G$ to denote only the visible part ($SU(5)$ or $SO(10)$) of the low-energy gauge group.

Notice that in terms of algebraic geometry, to get a bundle $V$ with structure group $G=SU(5)$ (resp. $G=SU(4)$), one must require that $\text{rank}(V) = 5$ (resp. $\text{rank} (V) = 4$) and $c_1(V) = 0$.

\subsubsection{GUT breaking}

We now have four-dimensional GUT theories, which we must break down to the Minimal Supersymmetric Standard Model (MSSM). However, in the context of heterotic compactifications we cannot use standard field theory mechanisms to break the gauge symmetry, since the massless spectrum of the heterotic string does not contain the required GUT Higgs field. We need a ``stringy'' mechanism. 

One approach that has proven fruitful is to break the GUT gauge group using a discrete Wilson line $F$ on $X$ (see \cite{GSW2}). This however requires that $F$ be the quotient of the fundamental group $\pi_1(X)$ by a normal subgroup. That is, we must require that $X$ be non-simply connected. This is the approach that we will follow in this paper.

For example, one could consider a Wilson line $F = \IZ_2$ to break the $SU(5)$ GUT group to the MSSM gauge group $SU(3) \times SU(2) \times U(1)$. In the case of $SO(10)$, one could consider $F= \IZ_6$ or $F=(\IZ_3)^2$ to break $SO(10)$ to the MSSM gauge group with an extra $U(1)$. Such breaking patterns can be obtained for many other finite groups $F$.

The moral here is that we want $\pi_1(X)$ to be as big as possible, since it provides more freedom for implementing realistic GUT breaking patterns.

\subsubsection{Massless particle spectrum}

The zero modes of the ten-dimensional Dirac operator on $X$ in the background with $V$ give rise to four-dimensional massless particles. Since $X$ is a Calabi-Yau threefold, the zero modes of the Dirac operators become zero modes of the Dolbeault operator on $X$ coupled to $V$. That is, four-dimensional massless particles can be represented by cohomology classes of certain bundles on $X$ associated to $V$.

One can work out the explicit correspondence between massless particles and cohomology groups by considering the decomposition of the adjoint representation of the visible $E_8$ under the embedding $G \times H$. However, for the purpose of this paper all that will be needed is the well known result that the net number of generations of massless particles is given by
\be
|h^1(X,V) - h^2(X,V)| = \frac{|c_3(V)|}{2},
\ee
where $h^\bullet(X,V)$ denotes the dimension of $H^\bullet(X,V)$, and we implicitly used the fact that for any stable vector bundle $V$,
\be
h^0(X,V) = h^3(X,V) = 0.
\ee

Therefore, to obtain three-generation models, we must require that
\be
c_3(V) = \pm 6.
\ee

\subsubsection{Tri-linear couplings}

Once we know the massless spectrum, we can ask what the tri-linear couplings in the low-energy Lagrangian look like. Since massless particles correspond to cohomology classes, it follows that tri-linear couplings can be computed through tri-linear products of cohomology groups. Specifically, consider the cohomology groups arising in the decomposition of the kernel of the Dirac operator, and form all possible triple cohomology products; those give the tri-linear couplings of the associated particles in the low-energy Lagrangian. We refer the reader to \cite{BCD} for more details.

\subsection{The heterotic challenge}

To conclude this section, let us combine the results in the form of an ``heterotic challenge", which we will try to solve in the remaining sections.

\begin{TWC}
Pick your favorite non-simply connected Calabi-Yau threefold $X$, and a K\"ahler form $\omega$ on $X$. The \emph{weak heterotic challenge} consists in constructing a rank $4$ or $5$ holomorphic vector bundle $V \to X$ such that
\begin{itemize}
\item $V$ is polystable with respect to $\omega$;
\item $c_1(V) = 0$;
\item $c_2(X) - c_2(V)$ is effective;
\item $c_3(V) = \pm 6$;
\item $\pi_1(X) = F$ where $F$ is a finite group that can be used to break the commutant $G$ of $H$ in $E_8$ --- where $H$ is the structure group of $V$ --- to the MSSM gauge group (perhaps with an extra $U(1)$ when $V$ has rank $4$).
\end{itemize}
\end{TWC}

We can also formulate a stronger challenge:

\begin{TSC}
Find a solution to the weak heterotic challenge, and show that:
\begin{itemize}
\item the full massless spectrum contains no exotic particles (apart from scalar moduli fields), by computing cohomology groups of vector bundles associated to $V$;
\item the tri-linear couplings are semi-realistic at tree level, by computing triple products of cohomology groups.
\end{itemize}
\end{TSC}

In this paper, we will restrict ourselves to the weak challenge, which we will try to solve on a new Calabi-Yau threefold $X$ that has not been studied from an heterotic string theory perspective yet. 

Let us note that so far the only solution to the strong heterotic challenge that has been found consists in the model of \cite{BD1,BCD}. We should also note that the strong challenge is of course not the end of the story, from a phenomenological point of view. For instance, we have not discussed supersymmetry breaking at all. Moreover, vacua which are solutions of the strong challenge will typically contain many massless scalar moduli fields, which need to be stabilized somehow. In short, more phenomenology is needed to make those vacua realistic. But this set of conditions, even in the weak form, turns out to be already very hard to solve.


\section{The Calabi-Yau threefold}

In the remainder of the paper, we will try to solve the weak heterotic challenge on a new Calabi-Yau threefold that has not been studied from an heterotic standpoint yet. Let us start by introducing in some detail the Calabi-Yau threefold, which was first described by Gross and Popescu \cite{GP} (see also \cite{GPa,BH}). Here we will only state the main features of the geometry; we refer the reader to the companion mathematical paper \cite{Ba} of the first author for proofs.

\subsection{The geometry}

\subsubsection{The cover}

Consider $\IP^7$ parameterized by $(x_0:\ldots:x_7)$. For any $(y_0:y_1:y_2) \in \IP^2$ in general position, consider the complete intersection $X_s$ of the four following quadrics:
\be
\begin{array}{l}
y_1y_3(x_0^2+x_4^2) - y_2^2(x_1x_7+x_3x_5)+(y_1^2+y_3^2)x_2x_6=0,\\
y_1y_3(x_1^2+x_5^2) - y_2^2(x_2x_0+x_4x_6)+(y_1^2+y_3^2)x_3x_7=0,\\
y_1y_3(x_2^2+x_6^2) - y_2^2(x_3x_1+x_5x_7)+(y_1^2+y_3^2)x_4x_0=0,\\
y_1y_3(x_3^2+x_7^2) - y_2^2(x_4x_2+x_6x_0)+(y_1^2+y_3^2)x_5x_1=0.\\
\end{array}
\ee
$X_s$ has $64$ double point singularities, and admits a free action of the group $F = \IZ_8 \times \IZ_8$.

While $X_s$ is singular, it can be shown that there exists a small resolution\footnote{In fact, there are two different small resolutions, which are related by a flop, but this fact will not be important for us \cite{BH}.} $\hX \to X_s$ of $X_s$ which is a smooth Calabi-Yau threefold. Moreover, $\hX$ has a fibration
\be
\hpi: \hX \to \IP^1,
\ee
whose generic fiber is an Abelian surface $A$ with $(1,8)$ polarization. It has topological data
\be
\chi(\hX) = 0, \qquad h^{1,1}(\hX) = h^{2,1}(\hX) = 2.
\ee

\subsubsection{The quotient}

The free action of $F= \IZ_8 \times \IZ_8$ on $X_s$ lifts to a free action on the smooth threefold $\hX$. Thus, the quotient
\be
X = \hX / F
\ee
is also smooth, and has fundamental group 
\be
\pi_1(X) = \IZ_8 \times \IZ_8. 
\ee
Its topological data is still
\be
\chi(X) = 0, \qquad h^{1,1}(X) = h^{2,1}(X) = 2,
\ee
 since both cohomology classes are invariant under the action of $F$. It turns out that $X$ is the Calabi-Yau threefold with the biggest fundamental group constructed so far.

The quotient map $\hX \to X$ agrees on the fibers with the isogeny $A \to A^\vee$, where $A^\vee$ is the Abelian surface dual to $A$.\footnote{\label{f:ab} The dual Abelian surface $A^\vee \simeq \text{Pic}^0(A)$ parameterizes isomorphism classes of flat line bundles on $A$. In contrast with the elliptic curve case, $A^\vee$ may not be isomorphic to $A$; rather, the map $A \to A^\vee$ is a surjective morphism with finite kernel (\emph{i.e.} an isogeny) --- we will come back to that when we construct bundles. For an Abelian surface with $(1,8)$ polarization, the kernel is $\IZ_8 \times \IZ_8$.} Hence, $X$ also has a fibration
\be
\pi: X \to \IP^1,
\ee 
whose generic fiber is an Abelian surface dual to the Abelian fiber of $\hX$. The Abelian fiber of $X$ also has $(1,8)$ polarization.

\subsubsection{Other threefolds}

\label{s:peaks}

Before we move on to the cohomology, it may be worth mentioning that it was shown by Borisov and Hua \cite{BH} that two other smooth Calabi-Yau threefolds with order $64$ (although non-Abelian) fundamental groups can be obtained as quotients of $\hX$. In fact, in \cite{Hu} Hua classified all smooth Calabi-Yau threefolds with non-trivial fundamental group that can be obtained as free quotients of small resolutions of singular complete intersections of four quadrics in $\IP^7$. Other related threefolds with non-trivial fundamental groups have also been explored in \cite{BD3,CD,DGS,Sa1,Sa2,Sa3}. All of those Calabi-Yau threefolds provide interesting new playgrounds to search for realistic heterotic compactifications. But for the purpose of this paper, we will restrict ourselves to the mother of them all, the Gross-Popescu threefold described above. It should however be possible to apply the techniques that we develop in this paper and in the companion paper \cite{Ba} to these other threefolds.

\subsection{The cohomology}
\label{s:coh}

To construct bundles on $X$, we need to understand the cohomology of $X$ and of its cover $\hX$. Let us start with $\hX$. We again refer the reader to \cite{Ba} for proofs of the statements.

\subsubsection{The cover}
\label{s:cover}

\begin{itemize}
\item
$H^2(\hX,\IZ)$ (modulo torsion) is generated by the pullback of the hyperplane section, $\hH$, and the class $\hA$ of the Abelian fiber of the fibration $\hpi$;
\item
$H^4(\hX, \IZ)$ (modulo torsion) is generated by the class of an exceptional curve $\he$ (which is a section of the fibration $\hpi$), and the class of a line $\hl$ in $X_s$ disjoint from the singular locus;
\item
$H^6(\hX, \IZ)$ is generated by the class of a point on $\hX$, $\qpt$.
\end{itemize}
The only non-zero intersection numbers are
\begin{gather}
\hH \cdot \hl = \qpt, \qquad \hA \cdot \he = \qpt, \qquad \hH^3 = 16 \qpt, \qquad \hH^2 \cdot \hA = 16 \qpt,\notag\\
\hH \cdot \hA = 16 \hl, \qquad \hH^2 = 16 \he + 16 \hl.
\end{gather}

\subsubsection{The quotient}
\label{s:quotient}

We now describe the cohomology of the quotient threefold $X$. 
\begin{itemize}
\item
$H^2(X,\IZ)$ (modulo torsion) is generated by the classes $\qH$ and $\qA$;
\item 
$H^4(X,\IZ)$ (modulo torsion) is generated by the classes $\qe$ and $\qE$, where $\qE$ is the class of a curve forming the singular locus of the singular fibers of the fibration $\pi$;
\item
$H^6(X,\IZ)$ is generated by the class of a point on $X$, $\qpt$.
\end{itemize}
 Denote by $\phi: \hX \to X$ the quotient map. The classes of $X$ are related to the classes of $\hX$ by
\begin{align}
&\phi^*(\qH) = 8 \hH,& \qquad &\phi_*(\hH) = 8 \qH,\notag\\
&\phi^*(\qA) = \hA,& \qquad &\phi_*(\hA) = 64 \qA,\notag\\
&\phi^*(\qe) = 64 \he,& \qquad &\phi_*(\he) =  \qe,\notag\\
&\phi^*(\qE) = 8 \hl,& \qquad &\phi_*(\hl) = 8 \qE,\notag\\
&\phi^*(\qpt) = 64 \qpt,& \qquad &\phi_*(\qpt) = 8 \qpt.
\end{align}

Using these relations we can compute the intersection ring on $X$. The result is that the only non-zero intersection numbers are (see \cite{Ba})
\begin{gather}
\qH \cdot \qE = \qpt, \qquad \qA \cdot \qe = \qpt, \qquad \qH^3 = 128 \qpt, \qquad \qH^2 \cdot \qA = 16 \qpt,\notag\\
\qH \cdot \qA = 16 \qE, \qquad \qH^2 = 16 \qe + 128 \qE.
\end{gather}
Note that the Chern classes of $X$ are
\be
c_1(X) = 0, \qquad c_2(X) = 8 \qE, \qquad c_3(X) = 0,
\ee



For the construction of bundles we need to know a little bit more about holomorphic curves in $\hX$ and $X$. Let $a \qe + b \qE$ be the class of a curve on $X$, and $a \he + b \hl$ be the class of a curve on $\hX$. For both threefolds, it is straightforward to show that the cone of effective curves is given by
\be
\{ a,b \in \IR~|~a,b \geq 0 \}. 
\ee



\subsubsection{The Abelian fibers}
\label{s:abelian}

In the process of constructing bundles on $X$, we will also need to understand the cohomology of the Abelian fibers of $\pi: X \to \IP^1$ in more detail. 

As mentioned previously, the generic fiber of $\pi$ is an Abelian surface $A$ with $(1,8)$ polarization. Its middle cohomology group $H^2(A,\IZ)$ is one-dimensional, and is generated by the polarization class, which we denote by $\qHA$.\footnote{From the threefold point of view, $\qHA$ corresponds to the restriction of the class $\qH$ to the Abelian fiber $\qA$.} Since the polarization class is of type $(1,8)$, its self-intersection is
\be
\qHA^2 = 16 \qpt.
\ee
If we denote by $a \qHA$ the class of a curve on $A$, the cone of effective curves is simply given by
\be
\{ a \in \IR~|~a \geq 0 \}.
\ee

While the generic fiber is an Abelian surface as above, not all fibers are of this type. The fibration $\pi$ contains $8$ singular fibers, which are all translation scrolls. Moreover, the fibration may contain Abelian surfaces which are ``non-generic", in the sense that they possess extra cohomology classes.

For instance, consider an Abelian surface $A$ which is the product of two elliptic curves, $A = E_1 \times E_2$. In this case, $H^2(A,\IZ)$ is three-dimensional, and is generated by the classes $\Eh$, $\Ev$ and $\Ed$, which correspond respectively to the class of the elliptic curve $E_1$, the class of $E_2$, and the class of the diagonal elliptic curve $E_d$. The non-zero intersection numbers are
\be\label{e:elliptic}
\Eh \cdot \Ev = 1, \qquad \Eh \cdot \Ed = 1, \qquad \Ev \cdot \Ed = 1.
\ee
If we denote by $x \Eh + y \Ev + z \Ed$ the class of a curve in $A$, it is easy to show that the cone of effective curves on $A$ is given by
\be
\{(x,y,z) \in \IR ~|~x y + x z + y z \geq 0, ~ x+ y + z \geq 0 \}.
\ee
Note however that this description is slightly redundant, since there is a transitive $SL(2,\IZ)$ action on the generators of the effective cone.

Finally, the following well known result will be useful in the following. Let $A$ be an Abelian surface, $[H]$ an ample class on $A$, and $[D]$ the class of a divisor. The Hodge Index Theorem implies that
\be\label{e:hodge}
[D]^2 \cdot [H]^2 \leq ([D] \cdot [H])^2.
\ee
This result follows from the fact that the intersection form on an Abelian surface is Lorentzian, which implies a reverse Cauchy-Schwarz inequality.


\section{Constructing bundles}
\label{s:bundles}

In order to solve the weak heterotic challenge we need to construct rank $4$ or $5$ stable holomorphic vector bundles $V \to X$ such that
\begin{itemize}
\item $c_1(V) = 0$;
\item $c_2(X) - c_2(V)$ is effective;
\item $c_3(V) = \pm 6$.
\end{itemize}
Since $c_2(X) = 8 \qE$, the second condition is equivalent to requiring that
\be
c_2(V) = (8-b) \qE - a \qe \qquad \text{for} \qquad a,b \geq 0.
\ee

In other words, we want to construct a stable holomorphic vector bundle $V \to X$ with Chern character
\be
\che(V) = r + (b-8) \qE + a \qe \pm 3 \qpt,
\ee
where $r \in \{4,5\}$ and $0 \leq a,b \in \IZ$. In the following we always assume that $a=0$, otherwise the Fourier-Mukai transform of $V$ has support on the whole threefold $\hX$, which renders the analysis rather difficult. We will come back to this in more detail in subsection \ref{s:fmv}.

Our main result is the following no-go theorem:

\begin{thm}\label{th}
Let $V$ be a stable holomorphic vector bundle on $X$ with respect to an ample class $[D] = [D_0] + k [A]$, where $k \gg 0$, $[D_0]$ is some fixed ample class on $X$, and $[A]$ is the class of the Abelian fiber.\footnote{\label{extra} Strictly speaking, we also need to assume that the restriction of $V$ to the singular fibers of $X \to \IP^1$ is semistable; see subsection \ref{s:singular}.} Then its Chern character cannot satisfy the phenomenological constraint
\be\label{e:chV}
\che(V) = r + (b-8) \qE \pm 3 \qpt,
\ee
where $r \in \{4,5\}$ and $0 \leq b \in \IZ$.
\end{thm}

The remainder of this section is devoted to a proof of this theorem. We first need to understand the Fourier-Mukai transform of vector bundles on $X$. Then we will argue that we can reduce $V$ to a simpler bundle on $X$ through a finite number of elementary transformations, and show that such elementary transformations imply that $V$ cannot have the desired Chern character.

\subsection{The Fourier-Mukai transform}

In order to prove the no-go theorem we will need to understand the Fourier-Mukai transform of vector bundles on $X$. We refer the interested reader to the books \cite{BBH,Huy} for a detailed description of the Fourier-Mukai transform in algebraic geometry.

The Fourier-Mukai transform of vector bundles on elliptically fibered Calabi-Yau threefolds (also sometimes called \emph{spectral cover construction}) is a well known topic that has been very useful for constructing phenomenologically interesting stable vector bundles, following the works \cite{FMW,Do,FMW2}. Here however the Calabi-Yau threefold that we are working with does not admit an elliptic fibration; rather, it possesses an Abelian surface fibration.

The mathematical theory of the Fourier-Mukai transform on Abelian surface fibrations is developed in the companion paper \cite{Ba}. Here, we only present a heuristic description of the construction, and state the main results that will be required for the purpose of the present paper.

Let us start by reviewing the main idea behind the spectral cover construction in the elliptic fibration case.

\subsubsection{Elliptic fibrations}

Start with an elliptic curve $E$. The space $\Pic^0(E)$ of isomorphism classes of flat (\emph{i.e.} degree $0$) line bundles on $E$ is an elliptic curve itself, $\Pic^0(E) \cong E^\vee$, which is called the \emph{dual} elliptic curve to $E$. It is well known that $E^\vee$ is isomorphic to $E$. Now consider a rank $n$ flat vector bundle $U \to E$ which splits as a direct sum of $n$ flat line bundles,
\be\label{e:sum}
U = L_1 \oplus \ldots \oplus L_n.
\ee
Each line bundle corresponds to a point on the dual elliptic curve $E^\vee$. Thus, the rank $n$ bundle $U$ can alternatively be described by a set of $n$ points on the dual elliptic curve $E^\vee$. Since $E^\vee \cong E$, we have a duality
\be
\{ U \to E\}  \qquad \leftrightarrow \qquad \{ n \text{ points on } E\}.
\ee
We say that the Fourier-Mukai transform of $U$ is the set of $n$ points. In fact, to be more precise we should say that the Fourier-Mukai transform of $U$ is the sum of $n$ skyscraper sheaves supported on those points.

Now fiber the construction over a two-dimensional base $B$. That is, consider a threefold $X$ which is an elliptic fibration over $B$, and a rank $n$ bundle $V \to X$ such that its restriction to every elliptic fiber is semistable and of degree $0$. The restriction of $V$ to a generic fiber then splits as a direct sum of $n$ flat line bundles. The bundle has an equivalent description in terms of the dual data $(C,L)$ on $X$, where the \emph{spectral surface} $C$ is a $n$-cover of the base $B$ (it intersects the generic elliptic fiber at $n$ points), and $L$ is a line bundle on $C$. We then have a duality
\be
\{ V \to X \} \qquad \leftrightarrow \qquad \{ (C,L) \text{ on } X\},
\ee
and we say that the Fourier-Mukai transform of $V$ is the spectral data $(C,L)$. We could also say that the Fourier-Mukai transform of $V$ is the sheaf $\cl = i_* L$, where $i: C \hookrightarrow X$ is the inclusion of the surface $C$ in $X$.

From a model building point of view, one of the main advantages of this construction is due to a theorem of Friedman, Morgan and Witten \cite{FMW2}. It says that if $C$ is an irreducible surface, then $V$ is necessarily stable in some region of the K\"ahler cone of $X$. Since proving stability of a bundle is in general rather difficult, this provides a straightforward way to construct stable bundles. However, the irreducilibity condition is only sufficient for stability, not necessary, and requiring irreducibility may be too strong a condition for phenomenological purposes; we may have to combine the spectral cover construction with other techniques, or consider reducible spectral covers.

\subsubsection{Abelian surface fibrations}

Now let us propose an analogous construction for Abelian surfaces. The mathematical details are provided in \cite{Ba}.

Start with an Abelian surface $A$. The space $\Pic^0(A)$ of isomorphism classes of flat line bundles on $A$ is again an Abelian surface, $\Pic^0(A) \cong A^\vee$, which is dual to $A$. As already mentioned in footnote \ref{f:ab}, a crucial point here is that $A$ is isomorphic to $A^\vee$ only if it is principally polarized. Otherwise, $A$ is only isogenous to $A^\vee$. In our case, since the polarization of $A$ is not principal, $A$ and $A^\vee$ are not isomorphic, only isogenous.

Consider a rank $n$ flat vector bundle $U \to A$ which splits as a direct sum of flat line bundles as in \eqref{e:sum}. Again, it has a dual description in terms of $n$ skyscraper sheaves on the dual Abelian surface $A^\vee$.

Now consider a Calabi-Yau threefold $X$ which is a fibration over $\IP^1$ whose fibers are Abelian surfaces. Let $\hX$ be the ``dual" threefold, in the sense that $\hX$ also has a fibration over $\IP^1$ where the fibers are dual to the fibers of $X$. Consider a rank $n$ bundle $V \to X$ such that its restriction to every Abelian fiber is semistable with vanishing Chern classes. The restriction of $V$ to a generic Abelian fiber then splits as a sum of $n$ flat line bundles. There is now an equivalent description in terms of the dual data $(C,L)$ on $\hX$, where the \emph{spectral curve} $C$ is a $n$-cover of the base $\IP^1$, and $L$ is a line bundle on $C$. We now have a duality
\be
\{ V \to X \} \qquad \leftrightarrow \qquad \{ (C,L) \text{ on } \hX\}.
\ee
What is important to note here is that when $A$ it not principally polarized, the bundle $V$ and its spectral data \emph{do not} live on the same threefold; rather, they live on ``dual" threefolds. This is in contrast with the principally polarized case, which includes elliptic fibrations.

This turns out to be a very interesting feature. Recall from the previous section that in the Gross-Popescu construction, the cover threefold $\hX$ and its quotient $X$ are precisely ``dual", in the sense described here, since their fibers are dual Abelian surfaces. Thus, we can construct a bundle on the quotient $X$ by specifying the spectral data on the cover $\hX$! Usually, in the ellipically fibered case, one constructs a bundle on the cover threefold, then shows invariance under the free action of the finite group, so that the bundle descends to a bundle on the quotient threefold. But here, it turns out that there is no invariance to prove, since we can construct a bundle \emph{directly} on the quotient $X$ by specifying its spectral data on the cover $\hX$.

Finally, note that in the Abelian surface case one can prove a stability result similar to the theorem of Friedman, Morgan and Witten; the proof is provided in \cite{Ba}. We must first define the notion of absolute irreducibility. Suppose that $V$ is a rank $n$ bundle described by the spectral data $(C,L)$. Then the bundles $\wedge^r V$, for $0 < r < n$, can also be described by spectral data $(C_r,L_r)$. We say that $V$ is \emph{absolutely irreducible} if $C_r$ is irreducible for all $0\leq r < n$, where $C_0 := C$. Then, the theorem of Bak states that if $V$ is absolutely irreducible, then it is stable in some region of the K\"ahler cone. Note however again that the condition of absolute irreducibility is only sufficient for stability, but not necessary.

\subsubsection{General vector bundles}

So far, we only described the Fourier-Mukai transform of vector bundles $V$ on $X$ such that their restriction to every Abelian fiber is semistable with vanishing Chern classes. But the Fourier-Mukai transform is much more general than that, and provides a functor between the derived categories of coherent sheaves on $X$ and $\hX$. In particular, to every vector bundle $V$ on $X$, we can associated a unique Fourier-Mukai dual sheaf (or complex of sheaves) on the dual threefold $\hX$.

The description of the Fourier-Mukai dual of a given vector bundle is in general quite complicated; it has been studied in detail in the companion paper \cite{Ba}. However, given a vector bundle $V$ on $X$, a simpler task is to extract the Chern character of its Fourier-Mukai dual sheaf (or complex of sheaves) $FM(V)$ on $\hX$. Indeed, the Fourier-Mukai transform provides a functor between the derived categories of coherent sheaves, and it also induces a compatible map on cohomology $fm : H^\bullet (X,\IQ) \to H^\bullet (\hX, \IQ)$, and an inverse map $fm^{-1}:  H^\bullet (\hX, \IQ) \to H^\bullet (X,\IQ)$. This cohomological map has been computed in \cite{Ba}; we reproduce here the result for convenience:\footnote{We note that the two first entries of those maps are the expected results, but have not been rigorously proved \cite{Ba}. In any case, they will not play a role in the following, only the last four entries will be necessary.}
\begin{align}
fm: \begin{pmatrix}\qX \\ \qH \\ \qA \\ \qe \\ \qE \\ \qpt \end{pmatrix} \to& \begin{pmatrix} \he + \qpt \\ - \hH + \frac{2}{3} \hA - 16 \hl + \frac{16}{3} \qpt \\ \qpt \\ \hhX + \hA \\ - \hl \\ \hA \end{pmatrix},\notag\\
fm^{-1}: \begin{pmatrix} \hhX \\ \hH \\ \hA \\ \he \\ \hl \\ \qpt \end{pmatrix} \to& \begin{pmatrix} \qe - \qpt \\ - \qH + \frac{16}{3} \qA + 16 \qE - \frac{2}{3} \qpt \\ \qpt \\ \qX - \qA \\ -\qE \\ \qA \end{pmatrix}.
\label{e:fm}
\end{align}
Using this cohomological map, given the Chern character of a vector bundle $V$ on $X$, we can extract the Chern character of its Fourier-Mukai dual on $\hX$.

\subsection{Proof of theorem \ref{th}: first step}
\label{s:step1}

\subsubsection{Fourier-Mukai dual of $V$}

\label{s:fmv}

As a first step in our proof of theorem \ref{th}, let us study the Fourier-Mukai dual of our desired vector bundles $V \to X$. We assume that $V$ is a stable holomorphic bundle on $X$ with Chern character
\be
\che(V) = r + (b-8) \qE + a \qe \pm 3 \qpt,
\ee
where $r \in \{4,5\}$ and $0 \leq a,b \in \IZ$. This is what is required for $V$ to satisfy the weak heterotic challenge.

Using the Fourier-Mukai cohomological map \eqref{e:fm}, we know that the sheaf (or complex of sheaves) $FM(V)$ on $\hX$ Fourier-Mukai dual to $V$ has Chern character
\be
\che(FM(V)) = a + (a \pm 3) \hA +(8-b) \hl + r \he + r \qpt.
\ee
The first thing to notice, as already mentioned in the beginning of section \ref{s:bundles}, is that if $a \neq 0$, the Fourier-Mukai dual sheaf (or complex of sheaves) $FM(V)$ has support on the whole threefold $\hX$, which is rather difficult to work with. But from a model building point of view, the point of using the Fourier-Mukai transform is to reduce the construction of bundles to a simpler dual standpoint. This is clearly not the case for $a \neq 0$. So in order to make the construction of physical bundles $V$ tractable, in the following we restrict ourselves to the case $a = 0$. 

In this case the Chern character of $V$ and its Fourier-Mukai dual read
\begin{align}
\che(V) =& r + (b-8) \qE \pm 3 \qpt, \notag\\
\che(FM(V)) =& \pm 3 \hA + (8-b) \hl + r \he + r \qpt.
\end{align}
From $\che(V)$, we obtain that the restriction of $V$ to every Abelian fiber is a rank $r$ bundle with vanishing Chern classes.

We assume that $V$ is stable with respect to an ample class $[D] = [D_0] + k [A]$, where $k \gg 0$, $[D_0]$ is some fixed ample class on $X$, and $[A]$ is the class of the Abelian fiber. By general principles,\footnote{Let us mention here that this implication is well known for elliptic fibrations, see for instance \cite{Fr,FMW2}. We claim that it also holds for Abelian surface fibrations.} this implies that the restriction of $V$ to almost all Abelian fibers $A$ is semistable with respect to the polarization class $[H_A]$ on $A$. In other words, there is only a finite number of Abelian fibers $A$ such that the restriction of $V$ to $A$ is unstable. 

We first claim that $V$ cannot be semistable on every Abelian fiber. Indeed, it is easy to compute that any vector bundle $V$ on $X$ whose restriction to all fibers is semistable with vanishing Chern classes must have $c_3(V) = 0$. This follows from the fact that for such vector bundle $V$, the Fourier-Mukai dual $FM(V)$ is supported on a spectral curve. The Chern character of $FM(V)$ can be computed as in subsection \ref{s:chernE}, and using the cohomological Fourier-Mukai transform one obtains that $c_3(V) = 0$; see \eqref{e:cheE}. Hence such vector bundles only yield non-chiral heterotic models. As a result, we know that the set of fibers on which $V$ is unstable must be non-empty.

From a Fourier-Mukai dual point of view, this implies that the support of the Fourier-Mukai dual $FM(V)$ of $V$ must be reducible; it must contain not only a spectral curve $s$ which is a $r$-cover of the $\IP^1$ base, but also extra vertical components. This also follows from the fact that the Chern character of $FM(V)$ includes a term $\pm 3 \hA$, since this term cannot be obtained from the Fourier-Mukai transform of a sheaf only supported on a spectral curve, see \eqref{e:fme}.

\subsubsection{Elementary transformations}

\label{s:et}

We have just seen that while the restriction of $V$ to \emph{all} fibers has vanishing Chern classes, $V$ restricts to a semistable bundle only on a \emph{generic} fiber. Indeed, there is a finite, non-zero, number of fibers over which $V$ is unstable.

It turns out that there is a canonical way to relate $V$ to an auxiliary bundle $E$ whose restriction to all fibers is semistable with vanishing Chern classes. We follow here the discussion in \cite{FMW2}, section 6.1 (see also \cite{Fr}), which can be easily generalized to the Abelian surface fibration case.

Since $V$ is semistable on a generic fiber, we can fix a bundle $E$ which restricts to a semistable bundle on all fibers and such that there exists a morphism $\phi: V \to E$ which is an isomorphism on a generic fiber. The map fails to be an isomorphism on a (not necessarily reduced) divisor $D$ in $X$, which is defined by $\det \phi$. $D$ is supported on a finite union of Abelian fibers, which correspond to the fibers over which $V$ is unstable. In fact, $D$ is the pullback of a divisor $d$ on $\IP^1$; let $\ell$ be the degree of $d$.

The main statement here, which parallels lemma 6.2 of \cite{FMW2}, is that $V$ can be ``transformed" into $E$ by a sequence of $\ell$ elementary transformations. Let us be a little more precise. Let
\be
Y = \{ y \in \IP^1 ~ | ~V_{A_y}\text{ is unstable} \},
\ee
where $V_{A_y}$ denotes the restriction of $V$ to the fiber $A_y$ above the point $y$ in the base. Choose one such point $y_1 \in Y$; $V_{A_{y_1}}$ must be unstable. Let $Q_1$ be the maximal destabilizing quotient sheaf for $V_{A_{y_1}}$.\footnote{The maximal destabilizing quotient sheaf $Q$ of an unstable bundle $E$ is defined by the quotient $E \to Q$ such that for every other quotients $E \to Q'$ we have that $\mu_H(Q') \geq \mu_H(Q)$, with $\mu_H(Q)$ the slope of $Q$ with respect to an ample class $H$, and equality only if the quotient factors through $E \to Q \to Q'$. In particular, the maximal destabilizing quotient sheaf $Q$ is necessarily semistable, since all its quotients $Q \to Q'$ must satisfy $\mu_H(Q') \geq \mu_H (Q)$.} There is then a short exact sequence
\be\label{e:et1}
0 \to V_1 \to V \to i_* Q_1 \to 0,
\ee
where $i$ is the inclusion of the fiber $A_{y_1}$ in $X$. Such short exact sequences are called \emph{elementary transformations of $V$} (or \emph{Hecke transforms}) in algebraic geometry. In the short exact sequence above, $Q_1$ is a torsion free sheaf, and $V_1$ is a torsion free reflexive sheaf.

The statement is then that there is a sequence of $\ell$ such elementary transformations which terminates to a vector bundle $E$ which is semistable on all fibers:
\begin{align}
0 \to V_1 \to V& \to i_* Q_1 \to 0,\notag\\
0 \to V_2 \to V_1& \to i_* Q_2 \to 0,\notag\\
\vdots& \notag\\
0 \to E \to V_{\ell-1}& \to i_* Q_{\ell} \to 0.
\label{e:seq}
\end{align}

\subsubsection{Chern characters}

\label{s:chernE}

We now have an explicit description of our bundle $V$ in terms of a sequence of elementary transformations \eqref{e:seq}. Let us now compute the Chern characters of the sheaves involved. 

First, we compute the Chern character of the auxiliary vector bundle $E$. Since $E$ is semistable with vanishing Chern classes on all fibers, we know that its Fourier-Mukai transform is a sheaf supported on a spectral curve which is a $r$-cover of $\IP^1$. Write $FM(E) = j_* L$, where $L$ is a sheaf supported on a curve $s$ in the class $r \he + m \hl$. Let $d = c_1(L)$. We know that $r \in \{4,5 \}$ by definition of $E$. Moreover, we know that $m \geq 0$ since $[s]$ must be effective. We can compute the Chern character of $FM(E)$ using Grothendieck-Riemann-Roch:
\begin{align}
\che(FM(E) ) =& j_* \( \che(L) \cdot \td(N_{s})^{-1} \) \notag\\
=& j_* (1+ d \qpt) \cdot (1 - (g-1) \qpt ) \notag\\
=& [s] + (d-g+1) \qpt \notag\\
=& r \he + m \hl + (d-g+1) \qpt.
\label{e:fme}
\end{align}
Here we used the normal sequence
\be
0 \to T_s \to T_{X|s} \to N_{s} \to 0
\ee
to compute the first Chern class of the normal bundle of $s$, $c_1(N_{s}) = (2g-2) \qpt$. Then, using the cohomological Fourier-Mukai transform \eqref{e:fm}, we obtain the Chern character of $E$:
\be\label{e:cheE}
\che(E) = r +(d-g+1-r) \qA - m \qE .
\ee

Second, recall from \eqref{e:chV} that we assumed
\be
\che(V) = r + (b-8) \qE \pm 3 \qpt .
\ee
Combining with the sequence of elementary transformations \eqref{e:seq}, we compute that
\begin{align}\label{e:chQ}
\sum_{i=1}^\ell \che(i_* Q_{i}) = & \che(V) - \che(E) \notag \\
=& (r +g-1-d) \qA + (b+m-8) \qE \pm 3 \qpt .
\end{align}

The second step of our proof consists in showing that elementary transformations \eqref{e:seq} with torsion free sheaves $Q_i$, for $i=1, \ldots, \ell$, supported on Abelian surfaces and satisfying \eqref{e:chQ} do not exist. This gives a contradiction, and shows that $V$ cannot have Chern character \eqref{e:chV}.

\subsection{Proof of theorem \ref{th}: second step}

In this section we show that elementary transformations \eqref{e:seq} with vector bundles $Q_i$ satisfying \eqref{e:chQ} do not exist. 

\subsubsection{Destabilizing short exact sequences}

Consider the elementary transformations in \eqref{e:seq},
\be
0 \to V_i \to V_{i-1} \to i_* Q_{i} \to 0
\ee
for $i=1, \ldots, \ell$, with $V := V_0$ and $E := V_{\ell}$. At each step, the torsion free sheaf $Q_i$ is defined as the maximal destabilizing quotient sheaf of the restriction of $V_{i-1}$ to $A_i$, which is the support of $Q_i$. We note here that by definition, the maximal destabilizing quotient sheaf $Q_i$ is semistable with respect to the polarization class on $A_i$. 

For clarity of notation, denote by $U_{i-1}$ the restriction of $V_{i-1}$ to $A_i$. The destabilizing short exact sequence reads
\be\label{e:senq}
0 \to N_i \to U_{i-1} \to Q_i \to 0,
\ee
where $N_i$ is a torsion free sheaf. Since $Q_i$ is a destabilizing quotient sheaf, and $U_{i-1}$ has vanishing Chern classes, the slope of $Q_i$ with respect to the polarization class $H_{A_i}$ on $A_i$ must be negative. That is,
\be
c_1(Q_i) \cdot [H_{A_i}]  \leq  0.
\ee
Correspondingly,
\be\label{e:slope}
c_1(N_i) \cdot [H_{A_i}]  \geq  0.
\ee

Moreover, it is easy to see that $N_i$ is related to the restriction of $V_i$ to $A_i$, which we denote by $W_i$, by an auxiliary short exact sequence,
\be\label{e:seaux}
0 \to Q_i \to W_i \to N_i \to 0.
\ee
For clarity, we now assume that all the Abelian surfaces $A_i$ are distinct. In other words, recalling the discussion in subsection \ref{s:et}, we assume that the divisor $D$ is reduced. We leave the proof in the non-reduced case to the reader. If all the $A_i$ are distinct, since the sequence of elementary transformations terminates to a vector bundle which is semistable on all fibers, we know that at each step $W_i$ is a semistable bundle with vanishing Chern classes.

\subsubsection{Chern characters}

Let us now compute what the requirements on the Chern characters of $Q_i$ and $N_i$ are. We assume first for simplicity that the Abelian surfaces $A_i$ are smooth; \emph{i.e.}, that the sheaves $Q_i$ do not live on some of the eight singular fibers of the fibration. The sheaves $Q_i$ have Chern characters
\be
\che(Q_i) = q_i + c_1(Q_i) + \frac{1}{2}(c_1(Q_i)^2 - 2 c_2(Q_i) ),
\ee
where $q_i$ denotes the rank of $Q_i$. Further, since the bundles $U_{i-1}$ have vanishing Chern classes, we get from \eqref{e:senq} that
\begin{align}
\che(N_i) =& n_i + c_1(N_i) + \frac{1}{2}(c_1(N_i)^2 - 2 c_2(N_i) )\notag\\
=&  r - q_i - c_1(Q_i) - \frac{1}{2}(c_1(Q_i)^2 - 2 c_2(Q_i) ),
\label{e:chN}
\end{align}
where we denoted by $n_i$ the rank of $N_i$.

Using Grothendieck-Riemann-Roch, we get that
\begin{align}
\che(i_* N_i) =& i_* \left( n_i + c_1(N_i) + \frac{1}{2}(c_1(N_i)^2 - 2 c_2(N_i) ) \right) \cdot \td(A_i)^{-1} \notag\\
 =& n_i \qA + i_* c_1(N_i) + \frac{1}{2}(c_1(N_i)^2 - 2 c_2(N_i) ),
\end{align}
and similarly for $Q_i$, where we used the fact that the Todd class of an Abelian surface $A_i$ is trivial.

Comparing with \eqref{e:chQ} and using \eqref{e:chN}, we obtain that the sheaves $N_i$ must satisfy:
\begin{align}
1.& \quad\sum_{i=1}^\ell (r-n_i) = r+g-1-d, \notag\\
2.&\quad \sum_{i=1}^\ell i_* c_1(N_i) = (8-b-m) \qE, \notag\\
3.& \quad \sum_{i=1}^\ell \che_2(N_i) = \frac{1}{2} \sum_{i=1}^\ell \( c_1(N_i)^2 - 2 c_2(N_i) \) = \pm 3 \qpt.
\label{e:condN}
\end{align}

Note that since $r+g-1-d$ can be an arbitrary integer so far, the first condition is harmless. 

The second condition can be reformulated directly on the Abelian surfaces $A_i$ on which the $N_i$'s are supported. Indeed, it is equivalent to the condition that 
\be
\sum_{i=1}^\ell c_1(N_i) \cdot [H_{A_i}] = 8-b-m,
\ee
where we recall from section \ref{s:abelian} that $[H_{A_i}]$ is the polarization class on the Abelian surface $A_i$. In particular, recall from section \ref{s:chernE} that $b \geq 0 $ and $m \geq 0$, so combining with \eqref{e:slope} we obtain that
\be\label{e:c11}
0 \leq \sum_{i=1}^\ell c_1(N_i) \cdot [H_{A_i}] \leq 8.
\ee

To summarize, the key point here is that the sequence of elementary transformations \eqref{e:seq} defines a set of $\ell$ torsion free sheaves $N_i$, supported on Abelian surfaces $A_i$, by the short exact sequences
\be\label{e:tt}
0 \to N_i \to U_{i-1} \to Q_i \to 0,
\ee 
where the $U_{i-1}$ have vanishing Chern classes. Moreover, the sheaves must satisfy
\be\label{e:ttt}
0 \leq \sum_{i=1}^\ell c_1(N_i) \cdot [H_{A_i}] \leq 8, \qquad \sum_{i=1}^\ell \che_2(N_i) = \pm 3 \qpt.
\ee
They also satisfy the auxiliary short exact sequences
\be
0 \to Q_i \to W_i \to N_i \to 0,
\ee 
where the $W_i$ are rank $r$ semistable vector bundles with vanishing Chern classes. 

We now that show that the two conditions \eqref{e:ttt} cannot be satisfied simultaneously. Our proof relies on successive applications of a reformulation of Bogomolov's inequality for semistable sheaves.



\begin{rem}
Recalling the discussion in section \ref{s:abelian}, naively one may expect that
\be\label{e:rem}
0 \leq \sum_{i=1}^\ell c_1(N_i) \cdot [H_{A_i}] \leq 8
\ee
implies that $c_1(N_i) = 0$ for all $i$, since on a generic smooth Abelian surface $c_1(N_i)$ must be a multiple of the polarization class $[H_{A_i}]$. However, this is only true if $A_i$ is a \emph{generic} Abelian surface. The fibration may contain non-generic smooth Abelian fibers with extra cohomology classes, such as $A_i$ being the product of two elliptic curves, in which case \eqref{e:rem} does not imply that $c_1(N_i) = 0$. Therefore, we cannot simply set $c_1(N_i) = 0$ in the following.
\end{rem}

\subsubsection{Bogomolov's inequality}

Let us start by recalling the well known Bogomolov's inequality for stable sheaves (see for instance chapter 9 of \cite{Fr}). In this section $A$ is always an Abelian surface.

\begin{lem}[Bogomolov's inequality]
Let $H$ be an ample divisor on $A$. Suppose that $V$ is a rank $v$ torsion free sheaf which is stable with respect to $H$. Then
\be
\che_2(V) = \frac{1}{2} c_1(V)^2 - c_2(V) \leq \frac{c_1(V)^2}{2 v} .
\ee 
\end{lem}

Combining this inequality with the Hodge Index Theorem \eqref{e:hodge}, we get
\be
\che_2(V) \leq \frac{c_1(V)^2}{2 v} \leq \frac{(c_1(V) \cdot H)^2}{2 v H^2}.
\ee
We now state and prove a simple lemma that extends this slightly weakened version of Bogomolov's inequality to semistable sheaves.

\begin{lem}
\label{ourlem}
Let $H$ be an ample divisor on $A$. Suppose that $V$ is a rank $v$ torsion free sheaf which is semistable with respect to $H$. Then
\be\label{e:bogo}
\che_2(V) \leq \frac{(c_1(V) \cdot H)^2}{2 v H^2} = \frac{v \mu_H(V)^2}{2 H^2},
\ee 
where
\be
\mu_H(V) = \frac{c_1(V) \cdot H}{v}
\ee
is the slope of $V$ with respect to $H$.
\end{lem}

\begin{proof}
If $V$ is semistable with respect to $H$, then there exists a filtration (see for instance exercise 16, p.117 in \cite{Fr}), known as the Jordan-H\"older filtration of $V$,
\be 
0 = E_0 \subset E_1 \subset \ldots \subset E_n = V,
\ee
such that the quotients $F_i = E_i / E_{i-1}$ are torsion free and $H$-stable for all $i$. Moreover, the quotients all have equal slope:
\be\label{e:sll}
\mu_H(F_1) = \ldots = \mu_H(F_n) = \mu_H(V).
\ee
So we have
\be\label{e:cc}
\che_2(V) = \sum_{i=1}^n \che_2(F_i) .
\ee
Each $F_i$ is stable, hence satisfies Bogomolov's inequality:
\be
\che_2(F_i) \leq \frac{c_1(F_i)^2}{2 f_i},
\ee
where $f_i$ denotes the rank of $F_i$.

Now by the Hodge Index Theorem (see \eqref{e:hodge}), since $H$ is ample, we have that
\be
c_1(F_i)^2 \leq \frac{(c_1(F_i) \cdot H)^2}{H^2}.
\ee
Thus,
\be
\che_2(F_i) \leq \frac{(c_1(F_i) \cdot H)^2}{2 f_i H^2} = \frac{f_i \mu_H(F_i)^2}{2 H^2}.
\ee
Using \eqref{e:cc}, we get
\be
\che_2(V) \leq \frac{1}{2 H^2} \sum_{i=1}^n f_i \mu_H(F_i)^2 = \frac{ v \mu_H(V)^2}{2 H^2},
\ee
where we used \eqref{e:sll} and the fact that $v = \sum_{i=1}^n f_i$.
\end{proof}

\subsubsection{Final step}

Let us now come back to the situation we are looking at. We have a sequence of $\ell$ short exact sequences on Abelian surfaces $A_i$:
\be\label{e:tt}
0 \to N_i \to U_{i-1} \to Q_i \to 0,
\ee 
and auxiliary short exact sequences
\be\label{e:QWN}
0 \to Q_i \to W_i \to N_i \to 0,
\ee 
where
\begin{itemize}
\item $U_{i-1}$ is a rank $r$ vector bundle with $c_1(U_{i-1})= 0 $ and $c_2(U_{i-1}) = 0$;
\item $W_i$ is a rank $r$ semistable vector bundle with $c_1(W_i) = 0$ and $c_2(W_i) = 0$;
\item $N_i$ is a rank $n_i < r$ torsion free sheaf with $0 \leq c_1(N_i) \cdot [H_{A_i}] \leq 8$;
\item $Q_i$ is a rank $q_i = r-n_i$ torsion free semistable sheaf with $c_1(Q_i) = - c_1(N_i)$ and $\che_2(Q_i) = - \che_2(N_i)$.
\end{itemize}
Note that as mentioned earlier, $Q_i$ is necessarily semistable, since it is defined as the maximal destabilizing quotient sheaf of $U_{i-1}$ in \eqref{e:senq}.

Let us now focus on one of these short exact sequences, and redefine for clarity $N:= N_i$, $U:= U_{i-1}$, $W := W_i$, $Q:= Q_i$, $A:= A_i$ and $H:= [H_{A_i}]$. We want to study the range of allowed $\che_2(N)$.

The first thing to notice is that since $Q$ is semistable, lemma \ref{ourlem} implies that
\be
\che_2(Q) \leq \frac{(c_1(Q) \cdot H)^2}{2 q H^2}.
\ee
Now $A$ has polarization $(1,8)$, which implies, as explained in section \ref{s:abelian}, that $H^2 = 16$. Moreover, since $c_1(Q) = -c_1(N)$, we know that
\be
-8 \leq c_1(Q) \cdot H \leq 0 \qquad \Rightarrow \qquad 0 \leq (c_1(Q) \cdot H)^2 \leq 64.
\ee
So we get the inequality
\be
\che_2(Q) \leq \frac{64}{32 q} = \frac{2}{q}.
\ee
But $q \in \IZ$ and $q >0$ by definition, and $\che_2(Q) \in \IZ$, so we obtain
\be\label{e:ch2Q}
\che_2(Q) \leq 2.
\ee

We now look at $N$. For any torsion free sheaf $N$, there is a filtration, known as the Harder-Narasimhan filtration (see for instance exercise 15, p. 112 in \cite{Fr}),
\be
0 = E_0 \subset E_1 \subset \ldots \subset E_n = N,
\ee
such that the quotients $F_i = E_i / E_{i-1}$ are torsion free and semistable and satisfy
\be\label{e:hn}
\mu_H(F_1) > \mu_H(F_2) > \ldots > \mu_H(F_n) .
\ee
At each step, the quotient $F_i$ is the maximal destabilizing quotient sheaf of $E_i$. Denote by $f_i$ the rank of $F_i$, and $e_i$ the rank of $E_i$. Note that by definition of the filtration, at the last level we have the short exact sequence
\begin{equation}
0 \to E_{n-1} \to N \to F_n \to 0,
\end{equation}
hence there is a surjective map $N \rightarrow F_n$. From \eqref{e:QWN}, there is also a surjective map $W \rightarrow N$, hence there is a surjective map $W \rightarrow F_n$. But this is a non-zero map between two semistable sheaves, hence we must have
\be
\mu_H(F_n) \geq \mu_H(W) = 0.
\ee
Thus \eqref{e:hn} can be strengthened to
\begin{equation}\label{e:hn2}
\mu_H(F_1) > \mu_H(F_2) > \ldots > \mu_H(F_n) \geq 0.
\end{equation}

Let us now prove that this implies that $\che_2(N) \leq 2$. From the filtration, we have that
\be\label{e:che2nn}
\che_2 (N) = \sum_{i=1}^n \che_2(F_i).
\ee
Each $F_i$ fits into a short exact sequence
\begin{equation}
0 \to E_{i-1} \to E_i \to F_i \to 0,
\end{equation}
where $F_i$ destabilizes $E_i$, so
\begin{equation}\label{eq:first}
0 \leq \mu_H(F_i) \leq \mu_H(E_i). 
\end{equation}
But by the short exact sequence we have
\begin{equation}
\mu_H(E_i)  = \frac{1}{e_i} \left(e_{i+1} \mu_H(E_{i+1} - f_{i+1} \mu_H(F_{i+1}) \right).
\end{equation}
Since
\begin{equation}
0 \leq \mu_H(F_{i+1}) \leq \mu_H(E_{i+1}),
\end{equation}
we obtain that
\begin{equation}
\frac{e_{i+1}-f_{i+1}}{e_i} \mu_H(E_{i+1}) = \mu_H(E_{i+1}) \leq \mu_H(E_i) \leq \frac{e_{i+1}}{e_i} \mu_H(E_{i+1}).
\end{equation}
Combining with \eqref{eq:first}, we get
\begin{equation}
0 \leq \mu_H(F_i) \leq \frac{e_{i+1}}{e_i} \mu_H(E_{i+1}).
\end{equation}
We keep going up to $E_n = N$ to obtain
\begin{equation}
0 \leq \mu_H(F_i) \leq \frac{n}{e_i} \mu_H(N).
\end{equation}
Finally, combining with our initial bound
\begin{equation}
0 \leq \mu_H(N) \leq \frac{8}{n},
\end{equation}
we get
\begin{equation}\label{eq:finalbound}
0 \leq \mu_H(F_i) \leq \frac{8}{e_i}.
\end{equation}
This bound is satisfied for each $F_i$.

Now each $F_i$ is semistable by definition of the filtration. Thus from lemma \ref{ourlem} we know that
\begin{equation}
\che_2(F_i) \leq \frac{f_i \mu_H(F_i)^2}{32}.
\end{equation}
Using \eqref{eq:finalbound}, we get that
\begin{equation}
\che_2(F_i) \leq 2 \frac{f_i}{e_i^2}.
\end{equation}
At level $i=1$, we have $E_1 = F_1$, so $e_1 = f_1$, and
\begin{equation}
\che_2(F_1) \leq 2.
\end{equation}
At each other level $i>1$, by definition of the filtration we must have $e_i > e_1 = 1$ and $f_i < e_i$. Recalling that $e_i$ and $f_i$ are integers, we get that,  for $i>1$,
\begin{equation}
\che_2(F_i) \leq 2 \frac{f_i}{e_i^2} < \frac{2}{e_i} \leq 1.
\end{equation}
That is,
\begin{equation}
\che_2(F_i) < 1.
\end{equation}
But $\che_2(F_i) \in \IZ$, hence, for $i>1$,
\begin{equation}
\che_2(F_i) \leq 0.
\end{equation}
Therefore, combining with \eqref{e:che2nn}, we get
\begin{equation}
\che_2(N) = \sum_{i=1}^n \che_2(F_i) \leq 2.
\end{equation}

What we have just shown is that
\begin{equation}
\che_2(N) \leq 2, \qquad \che_2(Q) \leq 2.
\end{equation}
But since $\che_2(N) = - \che_2(Q)$, this implies that
\be
-2 \leq \che_2(N) \leq 2.
\ee
Thus a single $N$ cannot lead to three generations as required by the second condition in \eqref{e:ttt}.

Now we are allowed to use more than one $N_i$ to satisfy the phenomenological conditions. However, the first condition of \eqref{e:ttt} tells us that
\begin{equation}
0 \leq \sum_{i=1}^\ell c_1(N_i) \cdot [H_{A_i}] \leq 8.
\end{equation}
By following the same argument as above, it is straightforward to show that if a given $N_i$ satisfies
\begin{equation}
0 \leq c_1(N_i) \cdot [H] \leq 5,
\end{equation}
then the conclusion of the argument is
\begin{equation}
\che_2(N_i) = 0.
\end{equation}
Hence for $\che_2(N_i) \neq 0$, it must satisfy
\begin{equation}
5 < c_1(N_i) \cdot [H_{A_i}] \leq 8.
\end{equation}
But clearly only one of the $N_i$ can satisfy this condition, since for all $N_i$'s we have that $c_1(N_i) \cdot [H_{A_i}] \geq 0$ by definition. Therefore, we conclude that
\begin{equation}
-2 \leq \sum_{i=1}^\ell \che_2(N_i) \leq 2,
\end{equation}
and the second condition of \eqref{e:ttt} cannot be satisfied. In other words, we can only get compactifications with at most two generations.

\subsubsection{Singular surfaces}

\label{s:singular}

We have shown that the existence of short exact sequences $0 \to Q_i \to W_i \to N_i \to 0$, with $0 \leq \sum_{i=1}^\ell c_1(N_i).[H_{A_i}] \leq 8$ and $W_i$ semistable with vanishing Chern classes implies that $-2 \leq \sum_{i=1}^\ell \che_2(N_i) \leq 2$, in contradiction with the three generation requirement.

However, strictly speaking we only analyzed the situation where the sheaves $N_i$ live on smooth Abelian fibers. To conclude the proof of theorem \ref{th} in full generality, we should also make sure that $N_i$ cannot live on some of the eight singular fibers of the Abelian surface fibration of $X$.

We believe that the argument above, with suitable modifications, also holds for a rank $n$ torsion free sheaf $N$ on a singular fiber. However, we should say that we do not have a rigorous proof of this, mostly due to subtelties related to sheaves on singular surfaces. Hence, to be rigorous we need to add in theorem \ref{th} the extra assumption that $V$ is semistable on the singular fibers of the fibration, which implies that the $A_i$ on which the elementary transforms in \eqref{e:seq}  are performed are smooth. This is the origin of footnote \ref{extra}, but we believe that this assumption could be safely removed.

To summarize, we have argued in this section that the sequence of elementary transformations \eqref{e:seq} that $V$ fits in produces a set of $\ell$ torsion free sheaves $N_i$ living on Abelian surfaces $A_i$. We showed that the sheaves $N_i$ cannot satisfy the conditions \eqref{e:condN}. This implies that the Chern character of $V$ cannot satisfy \eqref{e:chV}, which completes the proof of theorem \ref{th}.

\section{Conclusion}

In this paper we explored a new peak in the heterotic landscape. Namely, we tried to construct phenomenologically interesting heterotic compactifications on Gross-Popescu's Calabi-Yau threefold. This threefold has been singled out several times recently as an interesting place to look for realistic heterotic compactifications, since it possesses a large fundamental group. Our main result is a no-go theorem, which basically states that (under mild assumptions) there is no vector bundle on Gross-Popescu's threefold solving the weak heterotic challenge.

The negative conclusion that we reached in this paper is reminiscent of the analysis presented in \cite{BD2}. By studying compactifications on other threefolds with non-trivial fundamental groups, it was shown there that the anomaly cancellation condition and the stability conditions, which are both defining conditions for heterotic vacua to exist, are very difficult to satisfy simultaneously. However, relaxing one or the other leads to infinite families of ``frivolous" models, which may be understood as local models satisfying the numerical constraints but that cannot be consistently UV-completed in heterotic string theory. Here, it seems that a tension between the anomaly cancellation condition and the three-generation condition is at the core of our proof of the no-go theorem. Therefore, it seems again that requiring UV-completion of local models in heterotic string theory is very rigid, and throws away most of the realistic compactifications that one would naively construct from a local perspective. In fact, in the case of Gross-Popescu's threefold, requiring UV-completion eliminates \emph{all} phenomenologically interesting compactifications (under mild assumptions).

\subsection{Further avenues of research}

\begin{itemize}
\item
Although our no-go theorem is rather general, it relies on two distinct assumptions. It may be interesting to investigate whether dropping these assumptions leads to realistic models. 

First, we required that $a=0$, as explained in section \ref{s:fmv}. The reason being that for $a \neq 0$ the Fourier-Mukai transform of the bundle $V$ is supported on the whole threefold, which renders the analysis rather difficult. In other words, for $a \neq 0$, the bundle $V$ does not have vanishing second Chern class on a generic Abelian fiber. It would be interesting to see what happens for $a \neq 0$, but at the moment it is unclear to us how to approach this question, since our main tool, namely the Fourier-Mukai transform, cannot be used to provide a simpler dual description, at least at first sight. 

Second, as is standard in the spectral cover construction, we required stability of $V$ with respect to an ample class $[D] = [D_0] + k [A]$, where $k \gg 0$, $[D_0]$ is some fixed ample class on $X$, and $[A]$ is the class of the Abelian fiber. The reason for this choice is that it implies that the restriction of $V$ to a generic fiber must be semistable, which is one of the key ingredient in our proof of the no-go theorem. However, it may be worth investigating whether there exists interesting bundles which are stable with respect to ample classes far away from this region in the K\"ahler cone. Perhaps for such ample classes $V$ does not need to be semistable on a generic fiber, which could provide a way of evading our no-go theorem.

\item
As mentioned in section \ref{s:peaks}, it would be very interesting to explore other peaks in the Gross-Popescu range, such as those discovered in \cite{BH,DGS,Hu,Sa1,Sa2,Sa3}. The techniques developed in this paper and in the companion mathematical paper \cite{Ba} should be very useful in exploring these other peaks, which also possess Abelian surface fibrations. It would be interesting to see if no-go theorems can also be proved for these threefolds; in fact, it would be even better, from a phenomenological perspective, if one could use these threefolds to construct realistic models. In a similar vein, many new non-simply connected Calabi-Yau threefolds have also been constructed in \cite{BD3,CD}. It would be worth investigating whether phenomenologically interesting vector bundles can be constructed on these threefolds as well.

\item
On a different note, as alluded to in the introduction, it appears clear to us that the field of heterotic phenomenology is in a flagrant need of either a more algorithmic or a more intuitive approach to the construction of realistic bundles on Calabi-Yau threefolds.

On the algorithmic side, the first steps towards a computer-based search for realistic bundles have been laid out by Anderson, He and Lukas \cite{AHL,AHL2}. Their methods apply to a particular class of bundles, called \emph{monad bundles}. It would be very interesting to extend their approach to other types of bundles, or propose similar algorithmic techniques using methods such as the Fourier-Mukai transform or extensions.

On the intuitive side, there has been much progress recently on F-theory phenomenology (see for instance \cite{BHV,BHV2,DW,DW2}). In particular, in this context one obtains a direct geometric understanding of various phenomenological properties, such as the massless spectrum of the low-energy effective theory, the Yukawa couplings, etc. While the setup of heterotic theory is different from the F-theory setup, the geometries involved are quite similar, and it would be very interesting to see if one can extract from the geometric understanding that has been gained on the F-theory side a better intuitive approach to the construction of realistic bundles in heterotic string theory.

\end{itemize}

\end{document}